# Quantum *versus* Classical Descriptions of Spontaneous Emission in Nanophotonic Cavities

Jian-Hua Liang, Yue You, Xi-Hua Guan, Xiao-Jing Du, Jun He, and Zhong-Jian Yang*

*Hunan Key Laboratory of Nanophotonics and Devices, School of Physics, Central South University, Changsha 410083, China University, Changsha 410083, China*

[*zjyang@csu.edu.cn](mailto:zjyang@csu.edu.cn)

**ABSTRACT:** Here, we demonstrate that quantum and classical descriptions generally yield different results for the spontaneous emission in nanophotonic cavities. Starting from the quantized single-mode field in a general context of dispersive and lossy cavities, we derive the expression for emission rate enhancement as well as key relevant parameters such as mode volume and quality factor. For general nanophotonic cavities, this ratio of the quantum to the classical description is typically below unity and varies with the material dispersion properties, scattering-to-absorption ratio and morphology of the cavity. Notably, the two descriptions converge for lossless, non-dispersive dielectric cavities and for noble-metal plasmonic cavities with sufficiently low scattering losses.

The enhanced spontaneous emission of quantum emitters not only attracts attention in the field of fundamental physics but also finds numerous applications, including lasers[1,2], single-photon sources[3-5], single-emitter strong couplings[6-8], and so on. The reason why these aforementioned applications are feasible is that the environment can alter the emission rate of emitters—for instance, by incorporating plasmonic structures or photonic crystals[9-17].

Currently, there are mainly two types of theoretical approaches for describing the enhanced radiation of emitters near optical cavities. One is the classical approach, which is based on the classical Maxwell's equations, in which the emitter is typically treated as a classical electric dipole. In specific calculations, methods such as Green's function method, quasinormal modes (QNM), and numerical simulations are commonly used[18-30]. The other is the quantum approach, which is mainly based on Fermi's Golden Rule and utilizes the interaction between quantized fields and two-level systems[11,25,31-34]. Purcell first provided a quantum description in the microwave regime, thus, enhanced radiation is also referred to as the Purcell factor ($F_p$)[31,35]. With the development of nanophotonics, this concept has now been extended to the optical frequency regime, and the core parameter "volume" has originally evolved into the mode volume $V$ in nanophotonic cavities[19,36-39].

Digging deeper reveals a potential distinction between these two types of theoretical descriptions. For a resonant cavity with a given morphology, the classical approach exhibits a key characteristic: the electromagnetic field response and the corresponding emission results at a specific frequency only depend on the dielectric constant at that frequency[26,27], and are independent of the dielectric constant response at adjacent frequencies. On the other hand, the mode volume $V$—an important parameter in quantum descriptions—may be affected by dispersion[12,40], which means that the dielectric constant response at frequencies adjacent to a specific frequency also has an influence. The linewidth factor $\Delta\omega$ also appears in $F_p$ ($F_p \propto 1/(V\Delta\omega)$)[34]; although the factor $\Delta\omega$ is also affected by dispersion, the combination of the two ($V$ and $\Delta\omega$) cannot obviously cancel out the influence of dispersion formally. Thus, the quantum description of $F_p$ can formally include a dispersion term (Fig. 1). This motivates us to explore further the relationship between the two approaches.

In this work, starting from the quantized field of a single mode in a dispersive and lossy cavity, we derive the quantum description of $F_p$ as well as relevant parameters such as quality factor $Q$ and mode volume $V$. The derived $Q$ is generally different from the traditional one (denoted as $Q_T$). Importantly, for general nanophotonic cavities, the quantum and classical descriptions of emission rate enhancement exhibit discrepancies. The ratio of quantum description to classical one can be characterized by $Q_T/Q$, and it varies with the material dispersion, scattering-to-absorption ratio and morphology of a cavity. In experimentally feasible plasmonic systems, the discrepancy between the quantum and classical descriptions can even reach nearly a factor of ~0.6. On the other hand, the two descriptions converge in lossless and non-dispersive dielectric cavities or noble-metal plasmonic cavities with sufficiently low scattering losses.

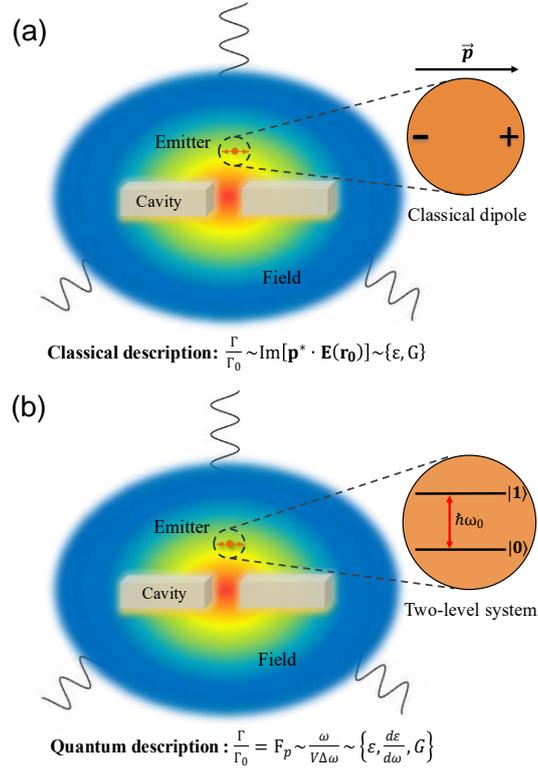

**Fig. 1.** Schematic diagram comparing the classical (a) and quantum (b) descriptions of spontaneous emission in nanophotonic cavities. In the classical description, the radiation enhancement at a given frequency $\omega$ is determined by the geometries (denoted by $G$) and the material's dielectric constant $\varepsilon$ at that frequency. In the quantum description, the radiation enhancement may also be influenced by dispersion $d\varepsilon/d\omega$.

**Quantum Description:**

The quantum description has a relatively mature framework[11,31,34,40-42]. First, according to Fermi's Golden Rule[34], the emission rate $\Gamma$ can be written as $\Gamma = \frac{2\pi}{\hbar^2}|g(\omega)|^2 \rho(\omega)$, where $g(\omega)$ is the interaction matrix element, $\rho(\omega)$ is the density of optical states. For a single mode, consistent with previous works, the cavity mode angular frequency is $\omega_c$, and its full width at half maximum (FWHM) is $\Delta\omega_c = 2\gamma_c$—here, $\gamma_c$ is the decay rate of the cavity mode. $\rho(\omega)$ can be expressed using a

Lorentzian line shape function $\rho(\omega) = \frac{1}{\pi\gamma_c}\frac{\gamma_c^2}{(\omega-\omega_c)^2+\gamma_c^2}$. The linewidth of the emitter is much smaller than that of the cavity resonant mode. Under resonance conditions, the enhanced emission rate (Purcell factor) can be derived as

$$F_p = \frac{\Gamma}{\Gamma_0} = \frac{3\omega\lambda_n^3}{4\pi^2 V \Delta\omega_c}, \qquad (1)$$

where $V$ is the mode volume parameter defined in field quantization. For a given optical cavity, $\gamma_c$ can be obtained via numerical simulation, while $V$ requires accounting for mode energy integration. To this end, we now consider a quasi-monochromatic (narrowband) optical field in an nanophotonic cavity composed of dispersive and lossy materials. The electric field satisfies

$$\boldsymbol{E}(\boldsymbol{r},t) = Re\big[\boldsymbol{E_1}(\boldsymbol{r},t)e^{-i\omega t}e^{-\gamma_c t}\big] = Re[\int\big(\boldsymbol{E_1}(\boldsymbol{r},\alpha)e^{-i(\omega+\alpha)t}e^{-\gamma_c t}\big)d\alpha], \quad (2)$$

where the time-dependent component of $\boldsymbol{E_1}(\boldsymbol{r},t)$ is a slowly varying term relative to $e^{-i\omega t}$, and $\alpha$ is the frequency component of the Fourier expansion of $\boldsymbol{E_1}(\boldsymbol{r},t)$. Meanwhile, the electric field has a $\gamma_c$-dependent decay term with time. Thus, the electric displacement $\boldsymbol{D}$ can be written as

$$\boldsymbol{D}(\boldsymbol{r},t) = \varepsilon\boldsymbol{E}(\boldsymbol{r},t) \approx Re\Big[\varepsilon_1\boldsymbol{E_1}(\boldsymbol{r},t)e^{-i\omega t}e^{-\gamma_c t} + i\frac{d\varepsilon_1}{d\omega}\frac{\partial \boldsymbol{E_1}(\boldsymbol{r},t)}{\partial t}e^{-i\omega t}e^{-\gamma_c t} +$$

$$i\varepsilon_2\boldsymbol{E_1}(\boldsymbol{r},t)e^{-i\omega t}e^{-\gamma_c t} - \frac{d\varepsilon_2}{d\omega}\frac{\partial \boldsymbol{E_1}(\boldsymbol{r},t)}{\partial t}e^{-i\omega t}e^{-\gamma_c t}\Big]. \qquad (3)$$

Here, $\varepsilon = \varepsilon_1 + i\varepsilon_2$ is the dielectric constant, where $\varepsilon_1$ and $\varepsilon_2$ are the real part and imaginary part, respectively. The energy density $u$ can be expressed as $\frac{\partial u}{\partial t} = \big(\boldsymbol{E}\cdot\frac{\partial \boldsymbol{D}}{\partial t} + \boldsymbol{H}\cdot\frac{\partial \boldsymbol{B}}{\partial t}\big)$. For the magnetic field term, its derivation is identical to that of the electric field. In typical nanophotonic cavities, the magnetic response is non-dispersive and $\mu = \mu_0$ (vacuum permeability), so the expression is further simplified. After some mathematical manipulations (see Supplementary Material), the

time-averaged total energy $\bar{U}$ is finally obtained

$$\bar{U} = \frac{1}{4}\int\left(|E_r|^2\left\{\varepsilon_1 + \omega\frac{d\varepsilon_1}{d\omega} + \gamma_c\frac{d\varepsilon_2}{d\omega}\right\} + |H_r|^2\mu_0\right)dv, \quad (4)$$

Here, $E_r$ and $H_r$ are the spatial distribution amplitudes of the electric field and magnetic field, respectively, and the integration range includes the cavity and the background environment. This expression is very similar to that proposed by Landau[43], except that we have incorporated a loss term here ($\gamma_c\frac{d\varepsilon_2}{d\omega}$). In practical systems, this term is very small; if this term is neglected, Eq. (4) becomes identical to Landau's formula.

Based on the standard derivation process for the interaction between quantized fields and emitters (see also Supplementary Material)[34,44], mode volume $V$ is given by

$$V = \frac{\bar{U}}{2\varepsilon_0 n^2|E_\mu|^2}, \quad (5)$$

where $E_\mu$ represents the component of the electric field along the oscillation direction at the position of the emitter, and $n$ is the refractive index at the position of the emitter. This expression is consistent with that presented in some previous works [13PRL]. Purcell's work was performed in closed, non-dispersive cavities, where $V$ refers to the cavity volume. Meanwhile, the linewidth factor $\omega/\Delta\omega_c$ at a specific frequency corresponds to the traditional expression for the quality factor ($Q = \omega/\Delta\omega_c$)[31]. However, in open cavities, the calculation of $V$ via integration often encounters an integral divergence issue, and integral truncation (equivalent to the effective spatial range of the mode) poses a challenge[19,37,45,46]. Additionally, the definition of the quality factor $Q$ itself demands careful consideration in dispersive cavities (see the discussion below).

To clarify the effective spatial range of the mode (i.e., the integral cutoff volume), we can consider the energy loss rate of the cavity. Based on Equation (4), the following expression can be derived (see Supplementary Material A)

$$-\overline{\frac{dU}{dt}} = -\frac{1}{4}\int\{|\boldsymbol{E_r}|^2\{-2\gamma_c\varepsilon_1 + 2\omega\varepsilon_2\} - 2\gamma_c|\boldsymbol{H_r}|^2\mu_0\}dv. \qquad (6)$$

For an optical cavity, its energy loss can be physically divided into two parts: absorption loss and scattering loss. Among these, the $\frac{1}{4}\int|\boldsymbol{E_r}|^2(2\omega\varepsilon_2)dv$ term corresponds to the absorption loss rate[23,47], so the remaining part should be the scattering loss. In the work of Ref.[47], only the static approximation was adopted, so the scattering loss was neglected. Physically, the ratio of the scattering loss rate to the absorption loss rate should be equal to the ratio of the scattering cross-section ($\sigma_{\text{scat}}$) to the absorption cross-section ($\sigma_{\text{abs}}$). Note that some numerical methods calculate the $\sigma_{\text{abs}}$ precisely by utilizing the proportional relationship between the absorption loss rate and the $\sigma_{\text{abs}}$. The absorption loss term generally only needs to be integrated over the entire cavity volume, as the exterior of the cavity is the background environment, which typically has no absorption loss. Furthermore, by calculating the $\sigma_{\text{scat}} / \sigma_{\text{abs}}$ of an optical resonant cavity, the scattering energy loss rate can be easily obtained. Thus, this determines the spatial integral cutoff range (the effective spatial range of the mode energy).

Figure 2 shows the effective volume cutoff range of a plasmonic nanosphere system, which strongly depends on the ratio $\sigma_{\text{scat}} / \sigma_{\text{abs}}$. The material follows a Drude model $\varepsilon = \varepsilon_\infty - \omega_p^2/(\omega^2 + i\gamma\omega)$ with $\varepsilon_\infty = 5$, $\omega_p = 1.2 \times 10^{16}\ rad/s$ and

$\gamma = 4 \times 10^{14}\ rad/s$. The cutoff volume can be ~$10^3$ times larger than the sphere's geometric volume and reaches the ~$\lambda^3$ scale. This is beyond the near-field region. Notably, even when scattering is negligible, a certain integral range outside the cavity is still required—because the integral of the scattering term inside metal is negative, and the exterior integral is needed to balance it out.

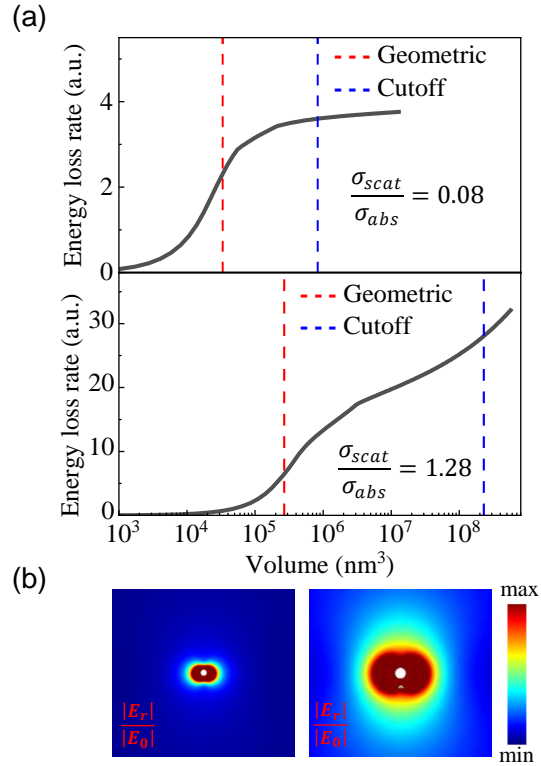

**Fig. 2.** The cutoff volumes (blue dashed lines) in plasmonic nanospheres under two cases of small and large scattering-to-absorption ratios. The red dashed lines indicate the geometric volume of the structures. The radius of the two spheres are 20 and 40 nm, respectively.

In Eq. (1), for the sake of rigor, we did not replace $\Delta\omega_c$ ($\gamma_c$) with the quality factor $Q$, because in a dispersive and lossy cavity, the $Q$ may not equal the traditional expression $Q_T = \omega/2\gamma_c$. According to the definition of the quality factor, $Q \equiv -\omega\overline{U}/(\overline{\frac{dU(t)}{dt}})$. Substituting Equations (4) and (6) into this expression yields

$$Q = \frac{\int(\omega|E_r|^2\{\varepsilon_1+\omega\frac{d\varepsilon_1}{d\omega}+\gamma_c\frac{d\varepsilon_2}{d\omega}\}+\omega|H_r|^2\mu_0)dv}{-\int\{|E_r|^2\{-2\gamma_c\varepsilon_1+2\omega\varepsilon_2\}-2\gamma_c|H_r|^2\mu_0\}dv}. \quad (7)$$

It is easy to see that for a non-dispersive dielectric cavity with no material loss, the above equation reduces to the traditional quality factor expression $Q = Q_T = \omega/2\gamma_c$. This corresponds to the case considered by Purcell and many other dielectric microcavities[31,48]. So in that scenario, writing the linewidth factor as $Q$ is valid (since $Q = Q_T$), and $F_p$ also becomes the familiar expression.

There is another special case, namely a plasmonic cavity where the scattering is negligible compared to absorption, in which Eq. (7) simplifies to $Q = \frac{\omega\frac{d\varepsilon_1}{d\omega}+\gamma_c\frac{d\varepsilon_2}{d\omega}}{2\varepsilon_2}$. For typical noble metal structures, $\omega\frac{d\varepsilon_1}{d\omega} \gg \gamma_c\frac{d\varepsilon_2}{d\omega}$, so the above expression can be approximated as

$$Q = \frac{\omega\frac{d\varepsilon_1}{d\omega}}{2\varepsilon_2}, \quad (8)$$

which is consistent with the result given in Ref.[47]. The condition adopted therein was the static approximation, which corresponds to the case where scattering is zero.

Further analysis reveals that in plasmonic systems, the calculation of $Q$ and $V$ does not actually require first determining the effective spatial mode range and then substituting this range into Eq. (7). Instead, for the effective spatial range, we can rewrite the energy loss rate as $\overline{\frac{dU(t)}{dt}} = \int\{\{-2\gamma_c\varepsilon_1 + 2\omega\varepsilon_2\}|E_r|^2 - 2\gamma_c\mu_0|H_r|^2\}dv_1 - \int\{-2\gamma_c\varepsilon_m|E_r|^2 - 2\gamma_c\mu_0|H_r|^2\}dv_2$, where $dv_1$ denotes the integration over the interior of the cavity material, and $dv_2$ denotes the integration over the external background medium. The expression for $\overline{U}$ can also be rewritten as the sum of internal and external integrals $\overline{U} = \int\left(\{\varepsilon_1 + \omega\frac{d\varepsilon_1}{d\omega} + \gamma_c\frac{d\varepsilon_2}{d\omega}\}|E_r|^2 + \mu_0|H_r|^2\right)dv_1 + \int\{2\varepsilon_m|E_r|^2 + 2\mu_0|H_r|^2\}dv_2$. It can be seen that the external integral in $\overline{U}$ differs from the external

integral in the loss rate term $\overline{\frac{dU(t)}{dt}}$ solely by a constant factor of $2\gamma_c$. And, the external integral in $\overline{\frac{dU(t)}{dt}}$ can be easily obtained by using the scattering-to-absorption ratio combined with the internal integral over the cavity material. This result is then substituted in the same way into the external integral term of $\overline{U}$. In this way, $Q$ and $V$ can be derived by calculating only the internal integral over the cavity material.

Furthermore, for $Q$ alone, its expression can be further optimized. The absorption component of the cavity can be isolated, and by letting the scattering-to-absorption ratio be $k$, the value of $Q$ under the condition of typical noble metal materials ($\omega \frac{d\varepsilon_1}{d\omega} \gg \gamma_c \frac{d\varepsilon_2}{d\omega}$) is

$$Q = Q_T \frac{k + \frac{\gamma_c \frac{d\varepsilon_1}{d\omega}}{-\varepsilon_2}}{k+1}. \tag{10}$$

Thus, one does not even need to integrate the near-field, and $Q / Q_T$ can be obtained solely using the far-field scattering and absorption spectra (from which $Q_T$ can also be derived) obtained via simulation, along with the material properties.

Generally, the $Q$ of a plasmonic structure deviates from $Q_T$. To highlight the effects of dispersion, $k$ ($\sigma_{\text{scat}}/\sigma_{\text{abs}}$), and structure morphology (which is related to the cavity mode decay rate $\gamma_c$) in general plasmonic cavities, we calculated the $Q / Q_T$ of plasmonic structures based on the Drude model (Fig. 3). This material model allows us to control and vary only one parameter at a time to observe its effect. It turns out that all three factors can indeed influence $Q / Q_T$. $Q / Q_T$ generally increases with $\sigma_{\text{scat}}/\sigma_{\text{abs}}$ and dispersion.

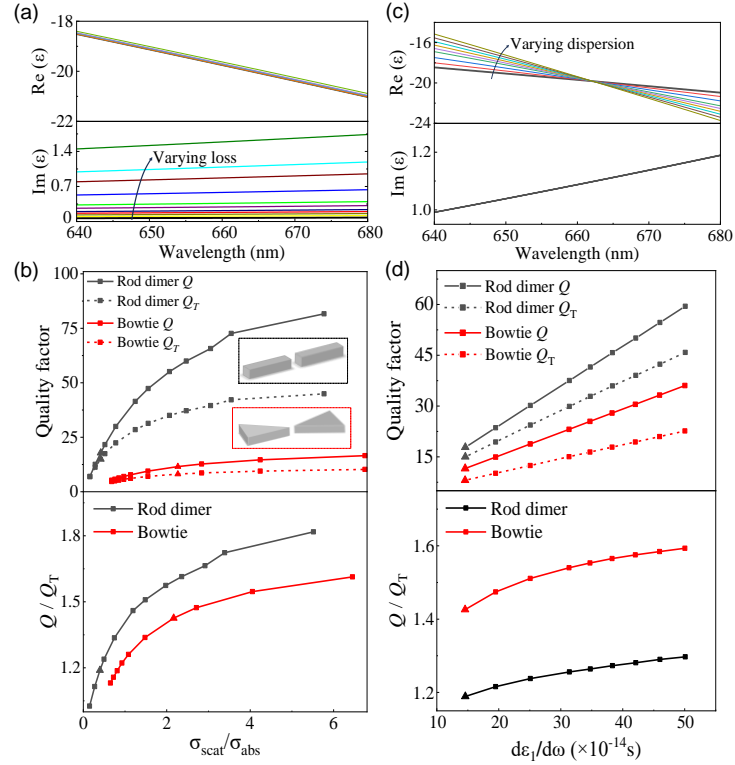

Fig. 3. (a) Drude models with different losses. (b) $Q$ and $Q_T$, as well as $Q/Q_T$, in rod dimers and bowtie structures as a function of $\sigma_{scat}/\sigma_{abs}$ (induced by different losses in (a)). Rod dimensions: $60\times14\times14$ nm$^3$; Bowtie triangle parameters: base length = 40 nm, side length = 87 nm, height = 30 nm. Gap size for both structures is 4 nm. (c,d) Drude models with different dispersion strengths from $14\times10^{-14}$s to $50\times10^{-14}$s. (c), and the resulting $Q$ and $Q_T$ as well as $Q/Q_T$ (d), in the aforementioned structures as a function of dispersion.

**Quantum *versus* Classical Descriptions:**

Discussions on classical emission rate enhancement are already abundant, including methods such as Green's function method, QNM, and direct numerical simulations[18-28] . Since we are focusing on the single-mode scenario, more analytical details can be obtained by combining the near-field coupling model between classical cavity fields and emitters. Relevant formulas have been provided in our previous similar work[49-51]. Under classical theory, the peak of enhanced emission rate (denoted as $F_p^C$) can be expressed as

$$F_p^C = \frac{6\pi c^2}{\omega^2} \frac{|E_\mu/E_0|^2}{\sigma_{ext}}, \qquad (11)$$

where $\sigma_{ext}$ is the extinction cross-section of the cavity, and $|E_\mu/E_0|$ is the component of the electric field enhancement along the polarization direction at the location where the electric dipole emitter is placed. Eq. (11) shows that $F_p^C$ is *not affected by dispersion*—for a nanophotonic cavity with a fixed geometry, namely the $F_p^C$ is determined solely by the material's dielectric constant corresponding to the resonant frequency, and frequencies in its vicinity exert no influence.

Notably, by substituting $\overline{U}$ into the numerator and denominator terms of Eq. (11) respectively, the following expression can be derived (see Supplementary Material)

$$F_p^C = \frac{3\lambda_n^3 Q}{4\pi^2 V}, \qquad (12)$$

where $Q$ and $V$ are the same as those defined previously (Eq. (7) and Eq. (5)). Although the classical result $F_p^C$ may formally contain $Q$ and $V$, the dispersion term in $\overline{U}$ is exactly canceled out. In contrast, the quantum result (denoted by $F_p^{QM}$) is generally related to dispersion. For the single-mode response of a dispersive and lossy resonant cavity, if one intends to express it using the quality factor, the quantum description formula Eq. (1) can be rewritten as

$$F_p^{QM} = \frac{3\lambda_n^3 Q_T}{4\pi^2 V}, \qquad (13)$$

where $F_p^{QM}$ denotes the quantum mechanical description of the Purcell factor. By comparing Eq. (12) and Eq. (13), the distinction between the quantum and classical descriptions can be expressed as

$$\frac{F_p^{QM}}{F_p^C} = \frac{Q_T}{Q}. \tag{14}$$

i.e., the ratio of $F_p^{QM}$ to $F_p^C$ is exactly equal to the ratio of $Q_T$ to $Q$.

It can be seen that for common dielectric microcavities, their $Q$ is inherently equal to $Q_T$. Thus, the classical and quantum results are identical, and no distinction is necessary. For plasmonic cavities, however, a distinction is generally required. From a formal perspective, some classical description expressions seem identical to quantum counterparts as the quality factor employed therein is also the traditional $Q_T$ [19]. In reality, these are still classical results because the integration range of $\bar{U}$ in $V$ is set to the near-field region therein. If the integration range of $\bar{U}$ in the $Q$ integral also adopts the near-field region, $Q$ will become $Q_T$ coincidentally. However, since $\bar{U}$ is canceled out in mode volume and qulity factor, the result remains consistent with that of $F_p^C$.

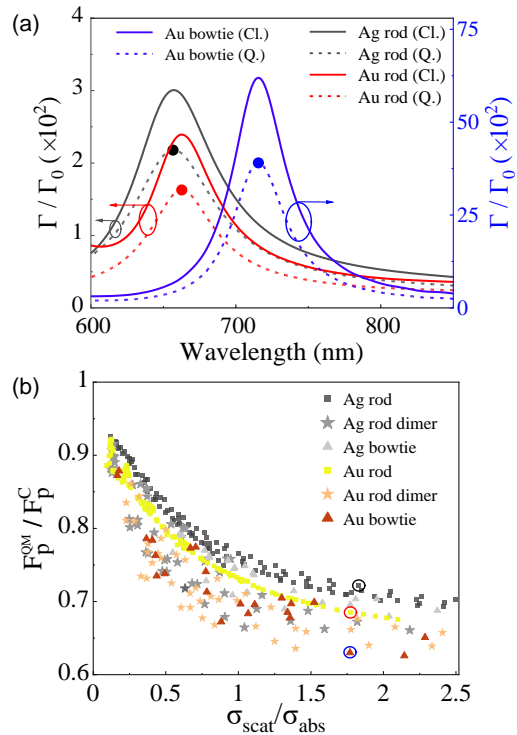

Fig. 4. (a) Classical and quantum results of emission-rate-enhancement spectra in practical plasmonic structures. The classical results are obtained directly via numerical simulation, while the quantum results are derived by first identifying the peaks (marked by solid points) and then fitting them using a Lorentzian lines. Dimensions of gold rods (silver rods): 110×40×40 nm$^3$ (120×34×34 nm$^3$); For the triangle of the bowtie structure: base length = 40 nm, side length = 80 nm, height = 30 nm, gap =10 nm. (b) Variations of the quantum-to-classical ratio with the scattering-to-absorption ratio for structures with different morphologies and resonance wavelengths. Among these, the three points corresponding to the cases in (a) are circled.

Since the quantum-to-classical ratio $F_p^{QM}/F_p^c$ is exactly equal to $Q_T/Q$, the previous discussion on the influencing factors of $Q_T/Q$ also applies to the quantum-to-classical ratio (Fig. 3). Next, let us examine the cases of practical plasmonic systems. Figure 4 shows a comparison of results for practical plasmonic systems with different scattering-to-absorption ratios, the structural morphology and resonance wavelengths (accounting for dispersion). For better material fittings in numerical simulations, the dielectric constants of gold and silver are adopted from CRC[52] and Palik,[53] respectively. It can be seen that in general plasmonic structures, the quantum results are generally smaller than the classical ones, and can even drop to ~0.6 times of the classical results. The scattering/absorption ratio has a significant impact on $F_p^{QM}/F_p^c$, while under the same scattering/absorption ratio, the structural morphology and dispersion also exert a certain degree of influence. When the scattering/absorption ratio approaches 0, the classical and quantum results converge. Typically, in plasmonic systems, the scattering/absorption ratio can be adjusted by modifying the structural volume[54].

Experimental verification of the quantum results can be conducted in plasmonic systems. However, the plasmonic system still faces challenges at present, mainly stemming from fabrication precision—especially physical fabrication methods. This is because the enhanced radiation is highly dependent on the distance between the emitter and the structure. Experimental errors caused by deviations in this distance can easily obscure the difference between the quantum and classical results. Chemical fabrication methods enable relatively precise control over the distance, but struggle to control factors such as the number of emitters. Nevertheless, with advances in nanofabrication technology, it is believed that verification will be achievable in the near future[55].

In conclusion, we have demonstrated that the quantum and classical descriptions of spontaneous emission in cavities generally exhibit discrepancies. Further studies indicate that the quality factor in dispersive and lossy cavities differs from the traditional definition, and the quantum-to-classical ratio $F_p^{QM} / F_p^c$ is exactly equal to $Q_\mathrm{T}/Q$. The calculation of $Q$ and $V$ is also relatively straightforward as it does not require actual integration over the effective mode integral cutoff range. Instead, it suffices to base the calculation on the cavity's scattering-to-absorption ratio combined with the internal cavity integral for $V$. As for the calculation of $Q$, integration is not even required. The quantum-to-classical ratio $F_p^{QM} / F_p^c$ varies not only with dispersion but also with the scattering/absorption ratio and morphology of the structure (accounting for $\gamma_c$). The ratio $F_p^{QM} / F_p^c$ is typically less than 1. In practical systems, the ratio $F_p^{QM} / F_p^c$ can even reach approximately a factor of ~0.6. Our work deepens the understanding of spontaneous emission in nanophotonic cavities and can have implications for enhancing light-matter interactions.

Supplementary Material for:

Quantum *versus* Classical Descriptions of Spontaneous Emission in Nanophotonic Cavities

Jian-Hua Liang, Yue You, Xi-Hua Guan, Xiao-Jing Du, Jun He, and Zhong-Jian Yang*

*Hunan Key Laboratory of Nanophotonics and Devices, School of Physics, Central South University, Changsha 410083, China University, Changsha 410083, China*

*zjyang@csu.edu.cn

**Part A: Electromagnetic energy**

The optical field in an cavity is assumed to be a quasi-monochromatic wave propagating in a dispersive and lossy cavity environment. For the electric field, we have

$$\boldsymbol{E}(\boldsymbol{r},t) = Re[\boldsymbol{E_1}(\boldsymbol{r},t)e^{-i\omega t}e^{-\gamma_c t}] = Re[\int (\boldsymbol{E_1}(\boldsymbol{r},\alpha)e^{-i(\omega+\alpha)t}e^{-\gamma_c t})d\alpha], \quad (A1)$$

Here, $\gamma_c$ denotes the decay rate of the cavity mode. According to $\boldsymbol{D}(r,t) = \varepsilon \boldsymbol{E}(r,t)$, a Taylor expansion can be performed as

$$\boldsymbol{D}(r,t) = Re\left[\int (\varepsilon_1(\omega+\alpha)\boldsymbol{E_1}(\boldsymbol{r},\alpha)e^{-i(\omega+\alpha)t}e^{-\gamma_c t})d\alpha + i\int (\varepsilon_2(\omega+\alpha)\boldsymbol{E_1}(\boldsymbol{r},\alpha)e^{-i(\omega+\alpha)t}e^{-\gamma_c t})d\alpha\right]$$

$$\approx$$

$$Re\left[\int \left((\varepsilon_1(\omega)+\alpha\frac{d\varepsilon_1(\omega)}{d\omega})\boldsymbol{E_1}(\boldsymbol{r},\alpha)e^{-i(\omega+\alpha)t}e^{-\gamma_c t}\right)d\alpha + i\int \left((\varepsilon_2(\omega)+\alpha\frac{d\varepsilon_2(\omega)}{d\omega})\boldsymbol{E_1}(\boldsymbol{r},\alpha)e^{-i(\omega+\alpha)t}e^{-\gamma_c t}\right)d\alpha\right]$$

$$= Re\left[\varepsilon_1\boldsymbol{E_1}(\boldsymbol{r},t)e^{-i\omega t}e^{-\gamma_c t} + i\frac{d\varepsilon_1}{d\omega}\frac{\partial \boldsymbol{E_1}(\boldsymbol{r},t)}{\partial t}e^{-i\omega t}e^{-\gamma_c t} + i\varepsilon_2\boldsymbol{E_1}(\boldsymbol{r},t)e^{-i\omega t}e^{-\gamma_c t} - \frac{d\varepsilon_2}{d\omega}\frac{\partial \boldsymbol{E_1}(\boldsymbol{r},t)}{\partial t}e^{-i\omega t}e^{-\gamma_c t}\right]. \quad (A2)$$

$\boldsymbol{E_1}(\boldsymbol{r},t)$ is a slowly varying function whose temporal variation rate is much smaller than the carrier frequency $\omega$, Therefore, the second-order derivative $\frac{\partial^2 \boldsymbol{E_1}}{\partial t^2}$ is negligible compared with

the first-order term, and we thus obtain

$$\frac{\partial \boldsymbol{D}}{\partial t} = Re\left[\frac{d(\varepsilon_1\omega)}{d\omega}\frac{\partial \boldsymbol{E}_1}{\partial t}e^{-i\omega t}e^{-\gamma_c t} - (\gamma_c + i\omega)\varepsilon_1 \boldsymbol{E}_1 e^{-i\omega t}e^{-\gamma_c t} - i\frac{d\varepsilon_1}{d\omega}\frac{\partial \boldsymbol{E}_1}{\partial t}\gamma_c e^{-i\omega t}e^{-\gamma_c t} + \right.$$

$$\left. i\frac{d(\varepsilon_2\omega)}{d\omega}\frac{\partial \boldsymbol{E}_1}{\partial t}e^{-i\omega t}e^{-\gamma_c t} + (\omega - i\gamma_c)\varepsilon_2 \boldsymbol{E}_1 e^{-i\omega t}e^{-\gamma_c t} + \frac{d\varepsilon_2}{d\omega}\frac{\partial \boldsymbol{E}_1}{\partial t}\gamma_c e^{-i\omega t}e^{-\gamma_c t}\right]. \tag{A3}$$

For convenience in calculation, we define $A$ as the real part of $\frac{\partial \boldsymbol{D}}{\partial t}$, Therefore

$$\boldsymbol{E} \cdot \frac{\partial \boldsymbol{D}}{\partial t} = Re[\boldsymbol{E}_1 e^{-i\omega t}e^{-\gamma t}] \cdot A$$

$$= \frac{1}{2}Re[\boldsymbol{E}_1^* e^{i\omega t}e^{-\gamma_c t}A] + \frac{1}{2}Re[\boldsymbol{E}_1 e^{-i\omega t}e^{-\gamma_c t}A]. \tag{A4}$$

Under time averaging, the term $\frac{1}{2}Re[\boldsymbol{E}_1 e^{-i\omega t}e^{-\gamma_c t}A]$ gives rise to higher-order oscillatory components, which are negligible compared with the first-order term, then

$$\boldsymbol{E} \cdot \frac{\partial \boldsymbol{D}}{\partial t} = \frac{1}{4}\frac{d(\varepsilon_1\omega)}{d\omega}\frac{d|\boldsymbol{E}_1|^2}{dt}e^{-2\gamma_c t} - \frac{1}{2}\gamma_c\varepsilon_1|\boldsymbol{E}_1|^2 e^{-2\gamma_c t} + \frac{1}{4}\gamma_c\frac{d\varepsilon_2}{d\omega}\frac{d|\boldsymbol{E}_1|^2}{dt}e^{-2\gamma_c t} + \frac{1}{2}\omega\varepsilon_2|\boldsymbol{E}_1|^2 e^{-2\gamma_c t}.$$

$$\tag{A5}$$

The energy density $u$ satisfies the following equation

$$\frac{\partial u}{\partial t} = \left(\boldsymbol{E} \cdot \frac{\partial \boldsymbol{D}}{\partial t} + \boldsymbol{H} \cdot \frac{\partial \boldsymbol{B}}{\partial t}\right). \tag{A6}$$

For the $\boldsymbol{E} \cdot \frac{\partial \boldsymbol{D}}{\partial t}$ term

$$\boldsymbol{E} \cdot \frac{\partial \boldsymbol{D}}{\partial t} = \frac{1}{4}\frac{d(\varepsilon_1\omega)}{d\omega}\frac{d|\boldsymbol{E}_1|^2}{dt}e^{-2\gamma_c t} - \frac{1}{2}\gamma_c\varepsilon_1|\boldsymbol{E}_1|^2 e^{-2\gamma_c t} + \frac{1}{4}\gamma_c\frac{d\varepsilon_2}{d\omega}\frac{d|\boldsymbol{E}_1|^2}{dt}e^{-2\gamma_c t} + \frac{1}{2}\omega\varepsilon_2|\boldsymbol{E}_1|^2 e^{-2\gamma_c t}.$$

$$\tag{A7}$$

For the magnetic field, we define $\boldsymbol{H}(\boldsymbol{r},t) = Re[\boldsymbol{H}_1(\boldsymbol{r},t)e^{-i\omega t}e^{-\gamma_c t}]$, and consider only the case of $\mu = \mu_0$. For the $\boldsymbol{H} \cdot \frac{\partial \boldsymbol{B}}{\partial t}$ term (which can be treated in a similar manner as the electric-field part above), we obtain

$$\boldsymbol{H} \cdot \frac{\partial \boldsymbol{B}}{\partial t} = \frac{1}{4}\frac{d(\mu\omega)}{d\omega}\frac{d|\boldsymbol{H}_1|^2}{dt}e^{-2\gamma_c t} - \frac{1}{2}\gamma_c\mu|\boldsymbol{H}_1|^2 e^{-2\gamma_c t} = \frac{1}{4}\mu_0\frac{d|\boldsymbol{H}_1|^2}{dt}e^{-2\gamma_c t} - \frac{1}{2}\gamma_c\mu_0|\boldsymbol{H}_1|^2 e^{-2\gamma_c t}.$$

$$\tag{A8}$$

Therefore, the electromagnetic energy $u$ is given as

$u(\boldsymbol{r},t) =$

$\frac{1}{4}\frac{d(\varepsilon_1\omega)}{d\omega}\int_0^t(\frac{d|E_1|^2}{dt}e^{-2\gamma_c t})dt - \frac{1}{2}\gamma_c\varepsilon_1\int_0^t|E_1|^2 e^{-2\gamma_c t}\,dt + \frac{\gamma_c}{4}\frac{d\varepsilon_2}{d\omega}\int_0^t\left(\frac{d|E_1|^2}{dt}e^{-2\gamma_c t}\right)dt +$

$\frac{1}{2}\omega\varepsilon_2\int_0^t|E_1|^2 e^{-2\gamma_c t}\,dt + \frac{1}{4}\mu_0\int_0^t(\frac{d|H_1|^2}{dt}e^{-2\gamma_c t})dt - \frac{1}{2}\gamma_c\mu_0\int_0^t|H_1|^2 e^{-2\gamma_c t}dt + u_0.$

(A9)

For clarity of interpretation, we assume that the envelope of $E_1(r,t)$ follows a Gaussian profile $E_1(r,t) = E_r(r)e^{-\frac{t^2}{2T^2}}$, $H_1(r,t) = H_r(r)e^{-\frac{t^2}{2T^2}}$. Therefore, Eq.(A9) can be simplified as follows

$u(r,t) =$

$\frac{1}{4}\frac{d(\varepsilon_1\omega)}{d\omega}e^{-2\gamma_c t}|E_r|^2 e^{-\frac{t^2}{T^2}} + \frac{\gamma_c}{2}\frac{d(\varepsilon_1\omega)}{d\omega}|E_r|^2\int_0^t(e^{-\frac{t^2}{T^2}}e^{-2\gamma_c t})\,dt -$

$\frac{\gamma_c\varepsilon_1}{2}|E_r|^2\int_0^t(e^{-\frac{t^2}{T^2}}e^{-2\gamma_c t})\,dt + \frac{\gamma_c}{4}\frac{d\varepsilon_2}{d\omega}|E_r|^2 e^{-2\gamma_c t}e^{-\frac{t^2}{T^2}} + \frac{\gamma_c^2}{2}\frac{d\varepsilon_2}{d\omega}|E_r|^2\int_0^t(e^{-\frac{t^2}{T^2}}e^{-2\gamma_c t})\,dt +$

$\frac{1}{2}\omega\varepsilon_2|E_r|^2\int_0^t(e^{-\frac{t^2}{T^2}}e^{-2\gamma_c t})\,dt + \frac{1}{4}\mu_0 e^{-2\gamma_c t}|H_r|^2 e^{-\frac{t^2}{T^2}} + \frac{1}{2}\mu_0\gamma_c|H_r|^2\int_0^t(e^{-\frac{t^2}{T^2}}e^{-2\gamma_c t})dt -$

$\frac{1}{2}\mu_0\gamma_c|H_r|^2\int_0^t(e^{-\frac{t^2}{T^2}}e^{-2\gamma_c t})\,dt + C,$  (A10)

where $C = -\frac{1}{4}\frac{d(\varepsilon_1\omega)}{d\omega}|E_r|^2 - \frac{\gamma_c}{4}\frac{d\varepsilon_2}{d\omega}|E_r|^2 - \frac{1}{4}\mu_0|H_r|^2 + u_0$. We can set

$$\int_0^t\left(e^{-\frac{t^2}{T^2}}e^{-2\gamma_c t}\right)dt = M.$$  (A11)

Then, it follows that

$u(r,t) =$

$\frac{1}{4}\frac{d(\varepsilon_1\omega)}{d\omega}e^{-2\gamma_c t}|E_r|^2 e^{-\frac{t^2}{T^2}} + \frac{\gamma_c}{2}\frac{d(\varepsilon_1\omega)}{d\omega}|E_r|^2 M - \frac{\gamma_c\varepsilon_1}{2}|E_r|^2 M + \frac{\gamma_c}{4}\frac{d\varepsilon_2}{d\omega}|E_r|^2 e^{-2\gamma_c t}e^{-\frac{t^2}{T^2}} +$

$\frac{\gamma_c^2}{2}\frac{d\varepsilon_2}{d\omega}|E_r|^2 M + \frac{1}{2}\omega\varepsilon_2|E_r|^2 M + C + \frac{1}{4}\mu_0|H_r|^2 e^{-2\gamma_c t}e^{-\frac{t^2}{T^2}}.$  (A12)

Since the envelope is slowly varying (weakly Gaussian wave packet), the term $e^{-\frac{t^2}{T^2}}$ can be approximated as 1, and under the weak-Gaussian condition $(t \ll T)$, we obtain

$$\int_0^t\left(e^{-\frac{t^2}{T^2}}e^{-2\gamma_c t}\right)dt = \frac{1}{2\gamma_c}(1 - e^{-2\gamma_c t}).$$  (A13)

$\overline{u(r,t)} = \frac{1}{t}\int_0^t u(r,t)dt$

$$= \frac{1}{4}\frac{d(\varepsilon_1\omega)}{d\omega}|E_r|^2\frac{M}{t} + \frac{\gamma_c}{2}\frac{d(\varepsilon_1\omega)}{d\omega}|E_r|^2\frac{1}{t}\int_0^t M\,dt - \frac{\gamma_c\varepsilon_1}{2}|E_r|^2\frac{1}{t}\int_0^t M\,dt + \frac{\gamma_c}{4}\frac{d\varepsilon_2}{d\omega}|E_r|^2\frac{M}{t}$$
$$+ \frac{\gamma_c^2}{2}\frac{d\varepsilon_2}{d\omega}|E_r|^2\frac{1}{t}\int_0^t M\,dt + \frac{1}{2}\omega\varepsilon_2|E_r|^2\frac{1}{t}\int_0^t M\,dt + \frac{1}{4}\mu_0|H_r|^2\frac{M}{t} + C$$
$$= \frac{|E_r|^2}{4t}\left\{\varepsilon_1\left(\frac{1}{2\gamma_c} - \frac{1}{2\gamma_c}e^{-2\gamma t}\right) + \omega\frac{d\varepsilon_1}{d\omega}t + \frac{d\varepsilon_2}{d\omega}\gamma_c t + \varepsilon_2\left(\frac{\omega}{\gamma_c}t - \frac{\omega}{2\gamma_c^2} + \frac{\omega}{2\gamma_c^2}e^{-2\gamma_c t}\right)\right\} + \frac{|H_r|^2}{4t}\left\{\mu_0\left(\frac{1}{2\gamma_c} - \frac{1}{2\gamma_c}e^{-2\gamma_c t}\right)\right\} + C. \tag{A14}$$

$$\overline{\frac{du(r,t)}{dt}} = \frac{1}{t}\int_0^t \frac{du(r,t)}{dt}dt = \frac{|E_r|^2}{t}\left\{\frac{1}{4}\frac{d(\varepsilon_1\omega)}{d\omega}(-2\gamma_c M) + \frac{\gamma_c}{2}\frac{d(\varepsilon_1\omega)}{d\omega}M - \frac{\gamma_c\varepsilon_1}{2}M + \frac{\gamma_c}{4}\frac{d\varepsilon_2}{d\omega}(-2\gamma_c M) + \frac{\gamma_c^2}{2}\frac{d\varepsilon_2}{d\omega}M + \frac{1}{2}\omega\varepsilon_2 M\right\} + \frac{|H_r|^2}{t}\left\{\frac{1}{4}\mu_0(-2\gamma_c M)\right\}$$
$$= \frac{1}{4t}\left\{|E_r|^2\left\{\left(\varepsilon_1 - \omega\varepsilon_2\frac{1}{\gamma_c}\right)(e^{-2\gamma_c t} - 1)\right\} + |H_r|^2\mu_0(e^{-2\gamma_c t} - 1)\right\}. \tag{A15}$$

Considering that $\bar{u}$ approaches zero as time tends to infinity, we obtain $C = 0$. Under the high-Q approximation, we have $(e^{-2\gamma_c t} - 1) \approx -2\gamma_c t$, and thus

$$\overline{U}(t) = \int \bar{u}\,dv = \frac{1}{4}\int\left(|E_r|^2\left\{\varepsilon_1 + \omega\frac{d\varepsilon_1}{d\omega} + \gamma_c\frac{d\varepsilon_2}{d\omega}\right\} + |H_r|^2\mu_0\right)dv, \tag{A16}$$

$$\overline{\frac{dU(t)}{dt}} = \frac{1}{4}\int\{|E_r|^2\{-2\gamma_c\varepsilon_1 + 2\omega\varepsilon_2\} + 2\gamma_c|H_r|^2\mu_0\}dv, \tag{A17}$$

$\overline{U}$ denotes the time-averaged total energy, and $\overline{dU(t)}/dt$ represents the energy loss rate of the cavity. According to the definition of the quality factor $Q$, we have

$$Q = \frac{\omega\overline{U}(t)}{-\overline{\frac{dU(t)}{dt}}} = \frac{\int(\omega|E_r|^2\{\varepsilon_1 + \omega\frac{d\varepsilon_1}{d\omega} + \gamma_c\frac{d\varepsilon_2}{d\omega}\} + \omega|H_r|^2\mu_0)dv}{-\int\{|E_r|^2\{-2\gamma_c\varepsilon_1 + 2\omega\varepsilon_2\} - 2\gamma_c|H_r|^2\mu_0\}dv}. \tag{A18}$$

## Part B: Quantum description of Purcell factor in a nanophtonic cavity

### 1. Quantization of the cavity field

The spontaneous emission transition rate is given by Fermi's golden rule as[S1]

$$\Gamma = \frac{2\pi}{\hbar^2}|g(\omega)|^2\rho(\omega), \tag{B1}$$

where $g(\omega)$ is the interaction matrix element and $\rho(\omega)$ denotes the photonic density of states. The single-mode electric field inside the cavity can be expressed as

$$E(r,t) = \text{Re}[aS(r)e^{-i\omega t}e^{-\gamma_c t}e^{-\frac{t^2}{T^2}}]$$

$$= \frac{a}{2}\mathbf{S}(r)e^{-i\omega_n t}e^{-\gamma_c t}e^{-\frac{t^2}{T^2}} + \frac{a}{2}\mathbf{S}(r)e^{i\omega_n t}e^{-\gamma_c t}e^{-\frac{t^2}{T^2}}, \tag{B2}$$

where $\mathbf{S}(r)$ denotes the normalized field distribution with a maximum value of 1, and $a$ represents the maximum field amplitude $E_{max}$. According to Eq. (A16), the time-averaged electromagnetic energy can be expressed in terms of $a$ as

$$\overline{U(t)} = \frac{1}{4}\int(|\mathbf{E_r}|^2\left\{\varepsilon_1 + \omega\frac{d\varepsilon_1}{d\omega} + \gamma_c\frac{d\varepsilon_2}{d\omega}\right\} + |\mathbf{H_r}|^2\mu_0)dv = c^2 a^2, \tag{B3}$$

where $c^2 = \frac{1}{4}\int(|\mathbf{S}(r)|^2\left\{\varepsilon_1 + \omega\frac{d\varepsilon_1}{d\omega} + \gamma_c\frac{d\varepsilon_2}{d\omega}\right\} + \frac{|\mathbf{H_r}|^2}{E_{max}^2}\mu_0)dv$. By quantizing the field as a harmonic oscillator, we define $a = \frac{\sqrt{\hbar\omega}}{c}\hat{a}$ and $a^* = \frac{\sqrt{\hbar\omega}}{c}\hat{a}^+$, where $\omega$ is the resonant frequency of the oscillator and $\hat{a}, \hat{a}^+$ are the photon annihilation and creation operators, respectively.

The quantized electric field is then expressed as

$$\mathbf{E}(r,t) = \frac{\sqrt{\hbar\omega}}{2c}\hat{a}\mathbf{S}(r)e^{-i\omega t}e^{-\gamma_c t}e^{-\frac{t^2}{T^2}} + \frac{\sqrt{\hbar\omega}}{2c}\hat{a}^+\mathbf{S}(r)e^{i\omega t}e^{-\gamma_c t}e^{-\frac{t^2}{T^2}}. \tag{B4}$$

The interaction Hamiltonian can be written as

$$\widehat{H}_{int} = -\widehat{\mathbf{P}}\cdot\widehat{\mathbf{E}}.$$

Under the second quantization and within the rotating-wave approximation, the interaction Hamiltonian can be expressed as

$$H_{int} = -\frac{\sqrt{\hbar\omega}}{2c}\boldsymbol{\mu_{10}}\cdot\mathbf{S}(r)(\hat{a}\hat{\sigma}_{10}e^{-i\omega t}e^{-\gamma_c t}e^{-\frac{t^2}{T^2}} + \hat{a}^+\hat{\sigma}_{01}e^{i\omega t}e^{-\gamma_c t}e^{-\frac{t^2}{T^2}})$$

$$= -g(\hat{a}\hat{\sigma}_{10}e^{-i\omega t}e^{-\gamma_c t}e^{-\frac{t^2}{T^2}} + \hat{a}^+\hat{\sigma}_{01}e^{i\omega t}e^{-\gamma_c t}e^{-\frac{t^2}{T^2}}), \tag{B5}$$

where $\boldsymbol{\mu_{10}}$ is the dipole transition matrix element between the two energy levels, $\hat{\sigma}_{10} = |1\rangle\langle 0|$ and $\hat{\sigma}_{01} = |0\rangle\langle 1|$ are the transition operators, and $g$ denotes the coupling strength.

$$g = \frac{\sqrt{\hbar\omega}}{2c}\boldsymbol{\mu_{10}}\cdot\mathbf{S}(r)$$

$$= \frac{\sqrt{\hbar\omega}}{2c}\boldsymbol{\mu_{10}}\cdot\frac{\mathbf{E}(r)}{E_{max}}$$

$$= \frac{\sqrt{\hbar\omega}}{2c}\boldsymbol{\mu_{10}}\cdot\frac{\mathbf{E_\mu}}{E_{max}} \tag{B6}$$

Here, $E_\mu$ denotes the electric field component along the oscillation direction at the emitter position. The mode volume of the cavity field is defined as[S2]

$$V = \frac{\int (|E_r|^2\{\varepsilon_1 + \omega\frac{d\varepsilon_1}{d\omega} + \gamma_c \frac{d\varepsilon_2}{d\omega}\} + |H_r|^2 \mu_0)dv}{2\varepsilon_0 n^2 |E_\mu|^2} = \frac{2c^2}{\varepsilon_0 n^2} \frac{E_{max}^2}{|E_\mu|^2}. \quad (B7)$$

The coupling matrix element $g$ can be expressed in terms of the field mode volume; thus, $g$ can be written as

$$g = \frac{1}{n}\sqrt{\frac{\hbar\omega}{2V\varepsilon_0}}\mu_{10} = \frac{1}{n}\sqrt{\frac{\hbar\omega\mu_{10}^2}{2V\varepsilon_0}}, \quad (B8)$$

where $\mu_{10} = |\boldsymbol{\mu_{10}}|$ represents the magnitude of the electric dipole moment.

## 2. Purcell factor

The emission spectrum is described by the spectral line-shape function $\rho(\omega)$. The nonresonant frequency broadening in Eq. (B5) is incorporated into the density of states $\rho(\omega)$. For a single mode, we have

$$\int_0^\infty \rho(\omega)d\omega = 1. \quad (B9)$$

Assuming the cavity resonance frequency to be $\omega_c$, the full width at half maximum to be $\Delta\omega_c = 2\gamma_c$, and the spectral function can be represented by a Lorentzian function as

$$\rho(\omega) = \frac{1}{\pi\gamma_c} \frac{\gamma_c^2}{(\omega-\omega_c)^2 + \gamma_c^2}. \quad (B10)$$

When the atomic transition frequency $\omega$ is in exact resonance with the cavity mode ($\omega = \omega_c$), the expression simplifies to

$$\rho(\omega) = \frac{1}{\pi\gamma_c}. \quad (B11)$$

For the cavity environment, substituting Eq. (B1), Eq. (B8), and Eq. (B11) yields the spontaneous emission rate inside the cavity

$$\Gamma = \frac{2\pi}{\hbar^2}|g(\omega)|^2 \rho(\omega) = \frac{2\omega\mu_{10}^2}{\hbar\varepsilon_0 nV\Delta\omega_c}. \quad (B14)$$

The spontaneous emission rate in free space is given by

$$\Gamma_0 = \frac{1}{\tau} = \frac{\mu_{10}^2 \omega^3}{3\pi n \varepsilon_0 \hbar c^3}. \tag{B15}$$

Therefore, the Purcell factor $F_p$ can be expressed as

$$F_p = \frac{\Gamma}{\Gamma_0} = \frac{3\omega(\lambda_n)^3}{4\pi^2 V \Delta\omega_c}. \tag{B16}$$

**Part C: Classical description of spontaneous emission enhancement**

In the classical case, the spontaneous emission enhancement at resonance can be expressed as

$$F_p^c = \frac{6\pi c^2}{\omega^2} \frac{|E_\mu/E_0|^2}{\sigma_{ext}}, \tag{C1}$$

where $\sigma_{ext}$ is the extinction cross section, $E_0$ denotes the electric field in free space, and $E_\mu$ represents the field in the cavity environment. We define $k = \sigma_{scat}/\sigma_{abs}$ as the ratio of scattering to absorption. According to the expression for the power dissipation rate, we have

$$P = (1+k) P_{abs}$$

$$= (1+k) \int \frac{1}{2} \omega \varepsilon_2 |E_r|^2 dv = \sigma_{ext} I$$

$$= \sigma_{ext} \frac{n}{2\mu_0 c} |E_0|^2 = \frac{1}{2} \sigma_{ext} \varepsilon c |E_0|^2. \tag{C2}$$

Therefore, we obtain

$$\sigma_{ext} = \frac{(1+k) \int \omega \varepsilon_2 |E_r|^2 dv}{\varepsilon c |E_0|^2}. \tag{C3}$$

Thus, Eq. (C1) can be simplified as

$$F_p^c = \frac{3\lambda_n^3}{4\pi^2} \frac{2\varepsilon_0 n^2 |E_\mu|^2}{\int 2\omega \varepsilon_2 |E_r|^2 dv} \frac{1}{(1+k)}. \tag{C4}$$

According to the definitions of $Q$ and $V$ in Eqs. (A18) and Eqs. (B7), we further obtain

$$F_p^c = \frac{3\lambda_n^3 Q}{4\pi^2 V}. \tag{C5}$$

Therefore, we demonstrate the equivalence of the two formulations of Eq. (C1) and Eq. (C5). In addition, by comparing with FDTD simulation results, we find that the spontaneous emission enhancement under all three conditions shows excellent agreement (Fig. S1).

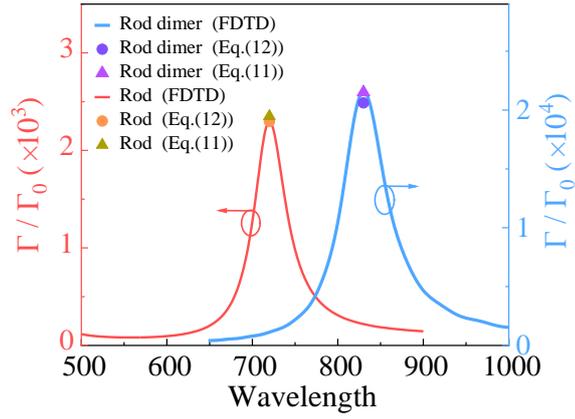

Fig.S1. Comparison of classical spontaneous emission enhancement under three calculation methods. Material: silver; nanorod dimensions: $100*20*20$ nm$^3$ (for a rod in both structures). The gap for the dimer is 10 nm. The solid line represents the spontaneous emission enhancement obtained from FDTD simulations, while the dots represent the values at the resonant peaks for the other two methods. Eqs. (C1) and (C5) are Eq. (12) and (11) in the mian text, respectively.